\documentclass[twocolumn,showpacs,preprintnumbers,amsmath,amssymb]{revtex4}
\usepackage{graphicx}

\begin{document}

\title{Positivity bounds on generalized parton distributions in impact
parameter representation}
\author{P.V. Pobylitsa}
\affiliation{
Institute for Theoretical Physics II, Ruhr University Bochum,
D-44780 Bochum, Germany\\
and Petersburg Nuclear Physics Institute, Gatchina, St. Petersburg, 188350,
Russia}
\pacs{12.38.Lg}

\begin{abstract}
New positivity bounds are derived for generalized (off-forward) parton
distributions using the impact parameter representation. These inequalities are
stable under the evolution to higher normalization points. The full set of
inequalities is infinite. Several particular cases are considered explicitly.
\end{abstract}

\maketitle

\section{General form of the positivity bounds on GPDs}

Generalized parton distributions (GPDs)
\cite{MRGDH-94,Radyushkin-96,Ji-97,CFS-97,Radyushkin-97,Radyushkin-review,GPV,BMK-2001} also known as
off-forward, skewed, nondiagonal etc. appear in the QCD description of
various hard processes e.g. deeply virtual Compton scattering and hard
exclusive meson
production. Our knowledge about GPDs is poor and any additional theoretical
information is of value. From this point of view the positivity bounds are
rather important.

GPDs are defined in terms of matrix elements
\[
G_{P_{1}i_{1},P_{2}i_{2}}^{\alpha _{1}\alpha _{2}}(x_{1},x_{2})=\langle
P_{2},i_{2}|\phi _{\alpha _{2}}^{\dagger }\left( x_{2}\right) \phi _{\alpha
_{1}}\left( x_{1}\right) |P_{1},i_{1}\rangle
\]
\begin{equation}
-\langle P_{2},i_{2}|P_{1},i_{1}\rangle \langle 0|\phi _{\alpha
_{2}}^{\dagger }\left( x_{2}\right) \phi _{\alpha _{1}}\left( x_{1}\right)
|0\rangle \,.  \label{G-def}
\end{equation}
Here $|P_{k},i_{k}\rangle $ is a hadron state (with momentum $P_{k}$ and
spin/isospin indices $i_{k}$) and the field $\phi _{\alpha _{k}}\left(
x_{k}\right) $ describes the annihilation of a parton with momentum fraction 
$x_{k}$ and with spin/isospin labeled by $\alpha _{k}$. Indices $\alpha_k$
also contain information about the type of the parton (quark or gluon).
The vacuum subtraction term in the RHS of Eq.~(\ref{G-def}) can be ignored
in practical applications of GPDs with $P_{1}\neq P_{2}$ but this term
is important in the derivation of the positivity constraints on GPDs.
Parton momentum fractions $x_i\equiv k_i^+/P^+$ are normalized with respect
to some external fixed scale $P^+$ and not to the hadron momenta $P_1$
or $P_2$.
Although notation (\ref{G-def}) for GPDs differs from the standard one,
we find the form (\ref{G-def}) rather convenient for the derivation of
positivity bounds and for the analysis of the interplay between the evolution
and the positivity properties.

The positivity of the norm in the Hilbert space of states
\begin{equation}
\left\| \sum\limits_{i\alpha }\int \frac{dP^{+}d^{2}P^{\perp
}dx}{2P^{+}(2\pi )^{3}}\theta (x)
f_{i\alpha}(x,P)
\phi _{\alpha}\left(
x\right) |P,i\rangle \right\| ^{2}\geq 0
\end{equation}
leads to the inequality
\[
\sum\limits_{i_{1}\alpha _{1}i_{2}\alpha _{2}}\int \frac{
dP_{1}^{+}d^{2}P_{1}^{\perp }dx_{1}}{2P_{1}^{+}(2\pi )^{3}}\int \frac{
dP_{2}^{+}d^{2}P_{2}^{\perp }dx_{2}}{2P_{2}^{+}(2\pi )^{3}}\theta
(x_{1})\theta (x_{2})
\]
\begin{equation}
\times \,f_{i_{2}\alpha _{2}}^{\ast }(x_{2},P_{2})f_{i_{1}\alpha
_{1}}(x_{1},P_{1})G_{P_{1}i_{1},P_{2}i_{2}}^{\alpha _{1}\alpha
_{2}}(x_{1},x_{2})\geq 0\,.  \label{starting-inequality}
\end{equation}
The integration over $x_{1,2}$ is restricted to the positive region
\begin{equation}
x_1>0\,,\quad x_2>0
\label{x-positive}
\end{equation}
where
the vacuum term vanishes in the RHS of Eq.~(\ref{G-def}) so that its subtraction
does not violate the positivity. In the case of the antiquark GPDs one should
consider the region $x_1,x_2<0$.

Strictly speaking the above expressions
have to be written more carefully:  in the case of the gauge
theories the $P$ exponent should be inserted between parton fields
and certain
conventions have to be chosen concerning the normalization of the momentum
fractions $x_{1},x_{2}$.

Various positivity inequalities for GPDs corresponding to specific choices
of the function $f_{i\alpha }(x,P)$ have been already discussed in
literature \cite{Martin-98, Radyushkin-99,
PST-99,Ji-98,DFJK-00,Burkardt-01,Pobylitsa-01,Pobylitsa-02,
Diehl-02,Burkardt-02-a,Burkardt-02-b}. The aim of this
paper is to analyze inequality (\ref{starting-inequality}) with arbitrary
functions $f_{i\alpha }(x,P)$.

\section{Positivity bounds and evolution}

It is well known that in the case of forward parton distributions the
probabilistic interpretation of the one-loop DGLAP evolution
\cite{GL-BLP,AP-77,Dokshitzer-77} leads to the stability of the positivity
properties under the one-loop evolution to higher normalization points. The
generalization of this property for some particular positivity bounds on
GPDs was considered in Ref. \cite{PST-99}. Let us show modifying the argument of
Ref. \cite{PST-99} that if at some
normalization point inequality (\ref{starting-inequality}) holds for all
functions $f_{i\alpha }(x,P)$ then after the one-loop evolution
\[
\mu \frac{\partial }{\partial \mu }
G_{P_{1}i_{1},P_{2}i_{2}}^{\alpha_{1}\alpha _{2}}(x_{1},x_{2};\mu )
=g^{2}(\mu )\int dy_{1}dy_{2}
\]
\begin{equation}
\times 
\sum\limits_{\beta_1\beta_2}
K_{\beta _{1}\beta _{2}}^{\alpha _{1}\alpha
_{2}}(x_{1},x_{2},y_{1},y_{2})G_{P_{1}i_{1},P_{2}i_{2}}^{\beta _{1}\beta
_{2}}(y_{1},y_{2};\mu )
\label{evolution}
\end{equation}
to a higher normalization point inequality (\ref{starting-inequality}) is
still valid for all $f_{i\alpha }(x,P)$.

Positivity bound (\ref{starting-inequality}) involves GPDs
$G_{P_{1}i_{1},P_{2}i_{2}}^{\alpha_{1}\alpha _{2}}(x_{1},x_{2};\mu )$
with positive $x_1,x_2>0$ (\ref{x-positive}). If $x_1,x_2>0$
then the one-loop evolution kernels
$K_{\beta _{1}\beta _{2}}^{\alpha _{1}\alpha
_{2}}(x_{1},x_{2},y_{1},y_{2})$
differ from zero only in the region
\begin{equation}
y_k \ge x_k \ge 0 \,.
\label{y-x-condition}
\end{equation}
This constraint has a simple physical meaning: a parton with momentum fraction
$y_k$ can emit a parton with momentum fraction $x_k$ only under the condition
(\ref{y-x-condition}).

The one-loop evolution kernels $K_{\beta _{1}\beta _{2}}^{\alpha _{1}\alpha
_{2}}(x_{1},x_{2},y_{1},y_{2})$ for GPDs can be interpreted as
perturbative
parton-in-parton GPDs. For parton-in-parton GPDs one can repeat the
derivation of the inequality (\ref{starting-inequality}) arriving at the
following inequality holding for any function $h_{\alpha \beta }(x,y)$
\[
\sum\limits_{i_{k}\alpha _{k}\beta_{k}}
\int\limits_0^\infty dx_1
\int\limits_{x_1}^\infty dy_1
\int\limits_0^\infty dx_2
\int\limits_{x_2}^\infty dy_2
h_{\alpha _{2}\beta _{2}}^{\ast
}(x_{2},y_{2})
\]
\begin{equation}
\times h_{\alpha _{1}\beta _{1}}(x_{1},y_{1})
K_{\beta _{1}\beta _{2}}^{\alpha _{1}\alpha
_{2}}(x_{1},x_{2},y_{1},y_{2})\geq 0\,.  
\label{kernel-positivity}
\end{equation}
The diagrammatic interpretation of this inequality is described in
Appendix \ref{positivity-kernels-appendix}.

Actually inequality (\ref{kernel-positivity}) holds only for functions
$h_{\alpha\beta}(x,y)$ vanishing at $x=y$ because of
the virtual terms proportional to $\delta
(x_{1}-y_{1})\delta (x_{2}-y_{2})$ which give a negative contribution to the
kernel $K_{\beta _{1}\beta _{2}}^{\alpha _{1}\alpha
_{2}}(x_{1},x_{2},y_{1},y_{2})$. If one ignores these terms then the
stability of the positivity bound (\ref{starting-inequality}) with respect
to the evolution (\ref{evolution}) upwards in $\mu$
is a consequence of the property
(\ref{kernel-positivity}) of the one-loop evolution kernel. The last step is to
notice that after the inclusion of the virtual terms proportional to $\delta
(x_{1}-y_{1})\delta (x_{2}-y_{2})$ in the evolution equation, the positivity
is still preserved under the evolution upwards. One can use the same
argument as in the case of the forward distributions: if at some ``critical
point'' $\mu $ inequality (\ref{starting-inequality}) is saturated and
becomes an equality for some function $f_{i\alpha }(x,P)$ then at this point
the virtual terms proportional to
$\delta (x_{1}-y_{1})\delta (x_{2}-y_{2})$ do
not contribute to the evolution of the LHS of inequality (\ref{starting-inequality}) so
that at higher $\mu $ the positivity is again restored by the positive part
of the evolution kernel. 
The details can be found in Appendix~\ref{positivity-evolution-appendix}.

\section{Positivity bounds in the impact parameter representation}

The positivity bound (\ref{starting-inequality}) contains a multidimensional
integral in the LHS and this form is not quite convenient for practical
applications. Turning to the impact parameter representation
\cite{Burkardt-01,Diehl-02,Burkardt-02-a,Burkardt-02-b} one can
simplify this inequality. Let us show how this can be done for quark GPDs.
The generalization for gluons is straightforward.

\begin{figure}
[ptb]
\begin{center}
\includegraphics[
height=1.228in,
width=3.0978in]
{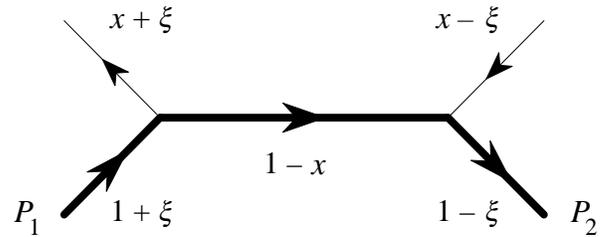}
\caption{Longitudinal momentum flow corresponding to variables $x,\xi$ [in
units of $(P_{1}^{+}+P_{2}^{+})/2$ ].}
\label{Fig-1}
\end{center}
\end{figure}

Let us introduce a light-cone vector $n$ and define the light-cone coordinates
so that $(nX)=X^{+}$ for any 4-vector $X$. Below for simplicity we shall
restrict our analysis to the case of ``unpolarized'' GPDs defined in terms of
the matrix elements of the operator $\bar{\psi}(n\gamma )\psi $
\[
\int \frac{d\lambda }{2\pi }\,\exp \left[ \frac{1}{2}i\lambda (k^+
_{1}+k^+ _{2})\right] 
\]
\[
\times\langle P_{2},\sigma _{2}|\bar{\psi}\left( -\frac{
\lambda n}{2}\right) (n\gamma )\psi \left( \frac{\lambda n}{2}\right)
|P_{1},\sigma _{1}\rangle
\]
\begin{equation}
=F_{\sigma _{2}\sigma _{1}}\left[ \frac{k^+ _{1}+k^+ _{2}}{
P_{1}^{+}+P_{2}^{+}},\frac{P_{1}^{+}-P_{2}^{+}}{P_{1}^{+}+P_{2}^{+}},\frac{
2(P_{1}^{+}P_{2}^{\perp }-P_{2}^{+}P_{1}^{\perp })}{P_{1}^{+}+P_{2}^{+}}
\right] \,.  \label{standard-GPD-F}
\end{equation}
The structure of the momentum dependence in
the RHS is fixed by the Lorentz invariance.
In order to see the relation between the arguments of function
$F_{\sigma _{2}\sigma _{1}}$ and the standard notation of X.~Ji \cite{Ji-98}
\begin{equation}
\Delta =P_{2}-P_{1}\,,\quad t=\Delta ^{2},\quad \xi =\frac{(nP_{1})-(nP_{2})
}{(nP_{2})+(nP_{1})}
\end{equation}
one can use the frame
where $P_{2}^{\perp }=-P_{1}^{\perp }\equiv \Delta ^{\perp }/2$. In this frame
the RHS of Eq.~(\ref{standard-GPD-F}) is simply
$F_{\sigma _{2}\sigma _{1}}(x,\xi ,\Delta ^{\perp })$.The longitudinal momentum flow corresponding to variables $x,\xi$
is shown in Fig.~\ref{Fig-1}.

Let us insert Eq. (\ref{standard-GPD-F}) into the positivity
condition (\ref{starting-inequality}) and turn to the impact parameter
representation
\[
\tilde{F}_{\sigma _{2}\sigma _{1}}(x,\xi ,b^{\perp })=\int \frac{d^{2}\Delta
^{\perp }}{(2\pi )^{2}}\exp \left[ i(\Delta ^{\perp }b^{\perp })\right]
\]
\begin{equation}
\times
F_{\sigma _{2}\sigma _{1}}(x,\xi ,\Delta ^{\perp })\,.  \label{F-Fourier}
\end{equation}
Using a factorized ansatz for the function $
f_{i_{1}\alpha _{1}}(x_{1},P_{1})$ one can reduce inequality
(\ref{starting-inequality}) to the following relatively simple form
(technical details can be found in Appendix~\ref{ineq-deriv-appendix})
\[
\sum\limits_{\sigma _{1}\sigma _{2}}\,
\int\limits_{-1}^{1}d\xi
\int\limits_{|\xi |}^{1}dx\frac{1}{(1-x)^{5}}p_{\sigma _{2}}^{\ast }\left( 
\frac{1-x}{1-\xi }\right) p_{\sigma _{1}}\left( \frac{1-x}{1+\xi }\right)
\]
\begin{equation}
\times \tilde{F}_{\sigma _{2}\sigma _{1}}\left( x,\xi ,\frac{1-x}{1-\xi ^{2}}
b^{\perp }\right) \geq 0\,.  \label{main-ineq}
\end{equation}
This inequality should be valid for any functions $p_\sigma$
(originating from the
factorized ansatz for $f_{i_{1}\alpha _{1}}(x_{1},P_{1})$) and for any value
of parameter $b^{\perp }$ if one wants the original inequality
(\ref{starting-inequality}) to hold for arbitrary $f_{i_{1}\alpha _{1}}(x_{1},P_{1})$.

For hadrons with spin 0 (e.g. pions) we define the GPD as follows
\[
q(x,\xi ,t)=\int \frac{d\lambda }{2\pi }\exp \left( i\lambda x\right)
\]
\begin{equation}
\times
\langle P_{2}|\bar{\psi}\left( -\frac{\lambda n}{2}\right) (n\gamma )\psi
\left( \frac{\lambda n}{2}\right) |P_{1}\rangle   \label{G-scalar}
\end{equation}
and for hadrons with spin 1/2 we use the standard notation of
X.~Ji \cite{Ji-98}
\[
\int \frac{d\lambda }{2\pi }\exp (i\lambda x)
\langle U(P_{2})|\bar{\psi}\left( -\frac{
\lambda n}{2}\right) (n\gamma )\psi \left( \frac{\lambda n}{2}\right)
|U(P_{1})\rangle 
\]
\[
=H(x,\xi ,t)\bar{U}(P_{2})(n\gamma )U(P_{1})
\]
\begin{equation}
+\frac{1}{2M}E(x,\xi ,t)\bar{U}
(P_{2})i\sigma ^{\mu \nu }n_{\mu }\Delta _{\nu }U(P_{1})
\label{H-E-def}
\end{equation}
Here we assume the normalization condition $n(P_{1}+P_{2})=2$. Note that we
could not impose this condition earlier since $P_{1}$ and $P_{2}$ were
integration variables in Eq. (\ref{starting-inequality}).
We also follow the standard convention that the transverse coordinates are
orthogonal to both $n$ and $P_{1}+P_{2}$. The Lorentz invariant squared
momentum transfer $t\equiv \Delta ^{2}<0$ can be expressed in terms of the
transverse part $|\Delta ^{\perp }|^{2}$
\begin{equation}
t= -(1-\xi ^{2})^{-1}\left( |\Delta ^{\perp }|^{2}+4\xi
^{2}M^{2}\right) \,.  \label{t-Delta-perp}
\end{equation}

In the case of spin-0 hadrons the function $F_{\sigma _{2}\sigma _{1}}$
appearing in inequality  (\ref{main-ineq})
can be expressed in terms of GPD $q$ (\ref{G-scalar}) as follows
\begin{equation}
F(x,\xi ,\Delta ^{\perp })=q\left( x,\xi ,-\frac{|\Delta ^{\perp }|^{2}+4\xi
^{2}M^{2}}{1-\xi ^{2}}\right)   \label{F-G-scalar}\,.
\end{equation}

For spin-1/2 hadrons we have
\[
F_{\sigma _{2}\sigma _{1}}(x,\xi ,\Delta ^{\perp })=\sqrt{1-\xi ^{2}}
\]
\begin{equation}
\times\left( 
\begin{array}{cc}
2H-\frac{2\xi ^{2}}{1-\xi ^{2}}E & \frac{-\Delta ^{1}+i\Delta ^{2}}{M(1-\xi
^{2})}E \\ 
\frac{\Delta ^{1}+i\Delta ^{2}}{M(1-\xi ^{2})}E & 2H-\frac{2\xi ^{2}}{1-\xi
^{2}}E
\end{array}
\right)_{\sigma _{2}\sigma _{1}}
 \label{F-H-E-nucleon} 
\end{equation}
where the arguments $x,\xi$ and $t$ (\ref{t-Delta-perp}) are assumed for $
H$ and $E$ in the RHS.

Inequality (\ref{main-ineq}) is our main result. Combining the expressions
for $F_{\sigma _{2}\sigma _{1}}$ in terms of GPDs
(\ref{F-G-scalar}), (\ref{F-H-E-nucleon})
with the impact parameter representation
(\ref{F-Fourier}) and inserting the result into inequality (\ref{main-ineq})
one can obtain the explicit form of the inequalities for GPDs $q,H,E$.

\section{Positivity bounds on nucleon GPDs}

From the practical point of view the most interesting case corresponds
to the positivity bounds on nucleon GPDs.
Let us rewrite our general inequality (\ref{main-ineq}) for spin-1/2
hadrons in terms of the standard notations for nucleon GPDs $H,E$
(\ref{H-E-def}).

Inserting Eqs.~(\ref{F-Fourier}) and (\ref{F-H-E-nucleon}) into
the general inequality (\ref{main-ineq}) we find
\[
\sum\limits_{\sigma _{1}\sigma _{2}}\,\int\limits_{-1}^{1}d\xi
\int\limits_{|\xi |}^{1}dx\frac{\sqrt{1-\xi ^{2}}}{(1-x)^{5}}p_{\sigma
_{2}}^{\ast }\left( \frac{1-x}{1-\xi }\right) p_{\sigma _{1}}\left( \frac{1-x
}{1+\xi }\right) 
\]
\[
\times \int \frac{d^{2}\Delta ^{\perp }}{(2\pi )^{2}}\exp \left[ i\frac{1-x}{
1-\xi ^{2}}(\Delta ^{\perp }b^{\perp })\right] 
\]
\begin{equation}
\times \left( 
\begin{array}{cc}
2H-\frac{2\xi ^{2}}{1-\xi ^{2}}E & \frac{-\Delta ^{1}+i\Delta ^{2}}{M(1-\xi
^{2})}E \\ 
\frac{\Delta ^{1}+i\Delta ^{2}}{M(1-\xi ^{2})}E & 2H-\frac{2\xi ^{2}}{1-\xi
^{2}}E
\end{array}
\right) _{\sigma _{2}\sigma _{1}}\geq 0  \label{nucl-ineq-1}\,.
\end{equation}
The integration over $\Delta ^{\perp }$ can be expressed in terms of the
integration over variable $t$ (\ref{t-Delta-perp}).
Since functions $p_{\sigma }(z)$ are arbitrary we can
rescale them: $p_{\sigma }(z)\rightarrow z^{3/2}p_{\sigma }(z)$. Then
inequality (\ref{nucl-ineq-1}) takes the form
\[
\int\limits_{-1}^{1}d\xi \int\limits_{|\xi |}^{1}\frac{dx}{(1-x)^{2}}
\int\limits_{-\infty }^{t_{0}}dt\,J_{0}\left( \frac{1-x}{\sqrt{1-\xi ^{2}}}
|b^{\perp }|\sqrt{t_{0}-t}\right) 
\]
\[
\times \left\{ \left( p_{+,-}^{\ast }p_{+,+}+p_{-,-}^{\ast }p_{-,+}\right)
\left( H-\frac{\xi ^{2}}{1-\xi ^{2}}E\right) \right. 
\]
\begin{equation}
\left. +\frac{p_{+,-}^{\ast }p_{-,+}+p_{-,-}^{\ast }p_{+,+}}{|b^{\perp
}|M(1-x)}\frac{\partial }{\partial t}\left[ (t_{0}-t)E\right] \right\} \geq 0\,.
\end{equation}
Here $J_{0}$ is the Bessel function, parameter
\begin{equation}
t_0=-\frac{4\xi^2M^2}{1-\xi^2}
\end{equation}
corresponds to the maximal kinematically allowed value of $t$ (\ref{t-Delta-perp}),
functions $H,E$ are taken at their
standard arguments $x,\xi ,t,$  and $p_{\sigma ,\pm }$ are the values of
arbitrary functions $p_{\sigma }(z)$ at points $z=(1-x)/(1\pm \xi )$
\begin{equation}
p_{\sigma ,\pm }=p_{\sigma }\left( \frac{1-x}{1\pm \xi }\right) \,.
\end{equation}

\section{Special cases}

The general positivity bound (\ref{main-ineq}) imposes rather serious
constraints on GPDs since it should hold for any functions $p_\sigma$
and for any values of the impact parameter $b^\perp$.
Let us show how choosing
various functions $p_\sigma$ one can reproduce most of the old positivity
bounds and obtain new interesting results.
For simplicity we
consider the case of spinless hadrons (the generalization for
spin-1/2 hadrons is straightforward).

1. Integrating inequality (\ref{main-ineq})
over $b^\perp$ with the weight $|b^{\perp }|^{2n}$ and replacing
$p(z)\to z^{n+2}p(z)$ we obtain the following inequality for GPD $q$
(\ref{F-G-scalar})
\[
\int\limits_{-1}^{1}d\xi \int\limits_{|\xi |}^{1}dx\frac{1}{(1-x)^{3}}
p^{\ast }\left( \frac{1-x}{1-\xi }\right) p\left( \frac{1-x}{1+\xi }\right)
\]
\begin{equation}
\times \left. 
\frac{\partial^n}{\partial t^n}
q\left( x,\xi ,t\right) \right| _{t=-4\xi ^{2}M^{2}/(1-\xi ^{2})}\ge0
\end{equation}
which should hold for any integer $n=0,1,2,\ldots $ and for any function $p$.

2. Taking $p(z)=\delta (z-c)$ in (\ref{main-ineq}) one arrives at

\begin{equation}
\int \frac{d^{2}\Delta^{\perp }}{
(2\pi )^{2}}\exp \left[ i(\Delta ^{\perp }b^{\perp })\right]
\left. q\left(x,0,t\right) \right| _{t=- |\Delta ^{\perp }|^{2}}\ge0\,.
\end{equation}
This coincides with the inequality derived earlier in Ref. \cite{Burkardt-01}.

3. Using function
\begin{equation}
p\left( z\right) =c_{1}\delta \left( z-\frac{1-x}{1+\xi }\right)
+c_{2}\delta \left( z-\frac{1-x}{1-\xi }\right)   \label{p-c-ansatz}
\end{equation}
in inequality (\ref{main-ineq}) we obtain
\[
|c_{1}|^{2}\left( \frac{1+\xi }{1-\xi }\right) ^{3}\tilde{F}\left( \frac{
x+\xi }{1+\xi },0,\frac{1-x}{1+\xi }b^{\perp }\right) 
\]
\[
+|c_{2}|^{2}\left( \frac{1-\xi }{1+\xi }\right) ^{3}\tilde{F}\left( \frac{
x-\xi }{1-\xi },0,\frac{1-x}{1-\xi }b^{\perp }\right) 
\]
\begin{equation}
+\frac{ c_{1}^{\ast }c_{2}+c_{2}^{\ast }c_{1} }{1-\xi ^{2}}
\tilde{F}\left( x,\xi ,\frac{1-x}{1-\xi ^{2}}b^{\perp }\right) \geq 0 \,.
\label{c-b-inequality}
\end{equation}
Taking into account that this inequality should hold for any $c_{1},c_{2}$
and rescaling $b^{\perp }\rightarrow (1-\xi^{2})(1-x)^{-1}b^{\perp }$ we
find
\[
(1-\xi ^{2})^{-1}\tilde{F}\left( x,\xi ,b^{\perp }\right) 
\]
\begin{equation}
\leq \sqrt{\tilde{F}(x_1,0,(1-\xi )b^{\perp })
\tilde{F}(x_2,0,(1+\xi )b^{\perp })
 }
\end{equation}
where 
\begin{equation}
x_{1,2}=(x\pm \xi )/(1\pm \xi )  \label{x-12-x-xi}\,.
\end{equation}
This inequality was derived in Ref. \cite{Diehl-02} (with a different
normalization used in the definitions of GPD $\tilde{F}$\ and of the impact
parameter $b^{\perp }$).

4. Taking $c_{k}=d_{k}\exp \left( -iP_k^{\perp }b^{\perp }\right) $ in
inequality (\ref{c-b-inequality}), integrating over $b^{\perp }$ and
optimizing the resulting inequality with respect to arbitrary coefficients $d_{k}$ we
obtain
\begin{equation}
q\left( x,\xi ,t\right) \leq \sqrt{q\left( x_{1}\right) q\left( x_{2}\right) 
}
\label{ineq-PST}
\end{equation}
where $x_1,x_2$ are given by Eq.~(\ref{x-12-x-xi}) and
$q\left( x_{1,2}\right) \equiv q(x_{1,2},0,0)$ is the usual forward parton
distribution.
This inequality was
obtained earlier in Ref. \cite{PST-99} (the authors of \cite{PST-99} use a different
normalization of GPD $q$).

5. Using the modification of the ansatz (\ref{p-c-ansatz})
corresponding to spin-1/2 hadrons one can derive from
the general inequality (\ref{main-ineq})
\[
\left. \left( H-\frac{\xi ^{2}}{1-\xi ^{2}}E\right) ^{2}+\left( \frac{E}{
2M(1-\xi ^{2})}|\Delta ^{\perp }|\right) ^{2}\right| _{x,\xi ,t}
\]
\begin{equation}
\leq \frac{
q(x_{1})q(x_{2})}{1-\xi ^{2}}\,.  \label{H-E-ineq-enahnced}
\end{equation}
Again relation (\ref{x-12-x-xi}) is assumed between the arguments $x,\xi$
in the LHS and the variables $x_{1,2}$ in the RHS. This inequality was derived earlier in
Ref. \cite{Pobylitsa-01}.

This short list of particular cases is only a small part of the bounds that
can be extracted from the general inequality (\ref{main-ineq}).

\section{Positivity bounds and renormalization}

GPDs are defined in terms of matrix elements (\ref{G-scalar}),
(\ref{H-E-def}) of parton fields 
separated by a light-cone interval. Similarly to the case of the {\em forward}
parton distributions (FPDs) this formal definition makes sense only 
in a combination with some renormalization procedure. Generally
speaking the renormalization includes subtractions which can violate
naive positivity bounds.

Working with the regularizations preserving the positivity of the norm
in the Hilbert space of states
(with all necessary comments concerning the color singlet sector, the
insertion of the $P$ exponent between parton fields etc.) one seems to be
on the safe ground, which gives an argument in favor of the validity of
the positivity bounds at high normalization points. 
On the other hand, in the case of FPDs
it is well known that only the cross sections associated with FPDs must
be positive whereas the naive positivity bounds may be violated  for FPDs
at low normalization points in nonphysical renormalization schemes
\cite{AFR-98}. One should keep in mind that
starting from the general inequality (\ref{starting-inequality})
one can reproduce the standard
positivity properties of FPDs.
Therefore a violation of the positivity properties of FPDs at low
normalization points would lead to the breakdown of the positivity bounds on
GPDs (this can be
directly seen in inequalities (\ref{ineq-PST}), (\ref{H-E-ineq-enahnced})
where the GPDs are constrained by FPDs).

In this paper we have shown that the validity of the positivity bounds on
GPDs at high normalization points is compatible with the one-loop evolution.
This self-consistency check is encouraging but certainly more serious
analysis is needed in order to clarify the status of inequality
(\ref{starting-inequality}) in the context of the renormalization.

\section{Conclusions}
Using the impact parameter representation we have derived positivity bound
(\ref{main-ineq})
on GPDs. This inequality should hold for any function $p_\sigma$
so that actually we deal with an infinite set of inequalities.
These positivity bounds impose certain constraints
on models of GPDs used in the phenomenological analysis of hard exclusive 
processes.

{\bf Acknowledgements.}
I appreciate discussions with A.~Belitsky, V.~Braun, J.C.~Collins, M.~Diehl,
L.~Frankfurt, D.S.~Hwang, X.~Ji, M.~Kirch, N.~Kivel, L.N.~Lipatov, A.~Manashov,
D.~M{\"{u}}ller, V.Yu.~Petrov, M.V.~Polyakov,
A.V.~Rad\-yu\-sh\-kin, M.~Strikman and O.~Teryaev.
This work was supported by DFG and BMBF.

\appendix

\section{Positivity properties of evolution kernels}
\label{positivity-kernels-appendix}
\begin{figure}
[ptb]
\begin{center}
\includegraphics[
height=1.1813in,
width=1.9233in
]
{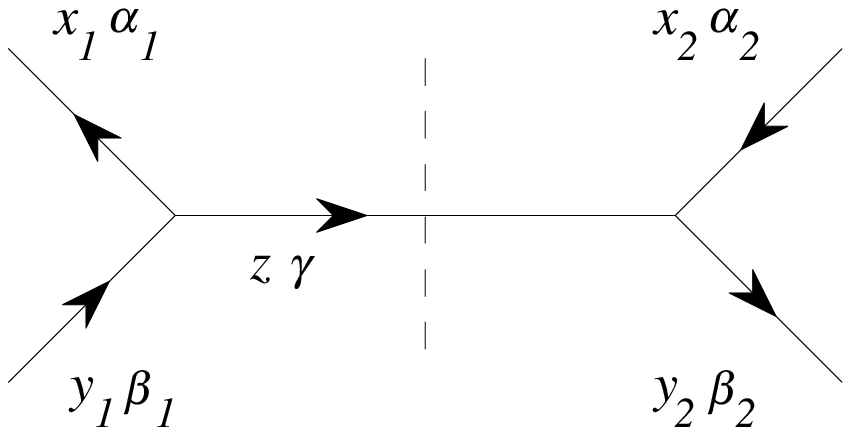}
\caption{Diagram representing the evolution kernel 
$K_{\beta_{1}\beta_{2}}^{\alpha_{1}\alpha_{2}}(x_{1},y_{1};x_{2},y_{2})$ at $0<x_{k}<y_{k}$.}
\label{Fig-2}
\end{center}
\end{figure}

In this Appendix we show how the positivity property (\ref{kernel-positivity})
of the evolution kernel $K_{\beta_{1}\beta_{2}}^{\alpha_{1}\alpha_{2}}
(x_{1},x_{2};y_{1},y_{2})$ can be seen in the direct diagrammatic calculation
of these kernels. Generally speaking, the leading order evolution kernel gets
contributions from several diagrams \cite{BFKL-85}
but in the region $0<x_{k}<y_{k}$ (\ref{y-x-condition}) which is interesting for us
this kernel is given by the single cut diagram of Fig.~\ref{Fig-2}. This
diagram leads to the following structure of the evolution kernel
\[
K_{\beta_{1}\beta_{2}}^{\alpha_{1}\alpha_{2}}(x_{1},x_{2};y_{1},y_{2}
)=\sum\limits_{\gamma}\delta(y_{1}-x_{1}-y_{2}+x_{2})
\]
\begin{equation}
\times V_{\alpha_{1}\beta
_{1}\gamma}(x_{1},y_{1})V_{\alpha_{2}\beta_{2}\gamma}^{\ast}(x_{2}
,y_{2})\,.
\label{K-alpha}
\end{equation}
In the framework of Ref.~\cite{BFKL-85} this
contribution naturally appears from the pole integration over the light-like
component of the momentum. In the case $x_{1},x_{2}>0$ the poles of the
propagators associated with $x_{1}$ and $x_{2}$ lie on the same side of the real
axis. Shifting the integration contour to the opposite side one can get a
nonzero contribution only from the pole of the propagator corresponding to the
intermediate parton in the diagram of Fig.~\ref{Fig-2}. Functions
$V_{\alpha_{k}\beta_{k}\gamma}(x_{k},y_{k})$ correspond to the two vertices of
the diagram of Fig.~\ref{Fig-2}. The kinematical and normalization factors
associated with the cut propagator are obviously positive (this positivity is
explicit in the light-cone gauge) and can be included into factors
$V_{\alpha_{k}\beta_{k}\gamma}(x_{k},y_{k})$. In the RHS of Eq. (\ref{K-alpha})
one sums over the polarization $\gamma$ of the intermediate parton.
Using the light-cone gauge and the techniques of Ref.
\cite{BFKL-85} it is easy to see that the summation over $\gamma$ is
restricted to the physical polarizations of the intermediate parton. Note that
we can rewrite (\ref{K-alpha}) in the form
\[
K_{\beta_{1}\beta_{2}}^{\alpha_{1}\alpha_{2}}(x_{1},x_{2};y_{1},y_{2})
\]
\begin{equation}
=\int\limits_{0}^{\infty}dz\sum\limits_{\gamma}W_{\alpha_{1}\beta_{1}
\gamma,z}(x_{1},y_{1})W_{\alpha_{2}\beta_{2}\gamma,z}^{\ast}(x_{2}
,y_{2})\label{K-beta-decomp}
\end{equation}
where
\begin{equation}
W_{\alpha\beta\gamma,z}(x,y)=\delta(y-x-z)V_{\alpha\beta\gamma}(x,y)\,.
\end{equation}
Obviously the form (\ref{K-beta-decomp}) of the evolution kernel $K$
automatically leads to the positivity property (\ref{kernel-positivity}).
Actually the general decomposition (\ref{K-beta-decomp})
is essentially equivalent to the inequality (\ref{kernel-positivity}). 
It is decomposition (\ref{K-beta-decomp}) that is needed for the proof of the stability of the positivity
bounds on GPDs with respect to the evolution to higher normalization points.

In order to illustrate the above general formulas with an explicit example let
us consider the evolution kernel for the ``helicity-independent'' quark GPDs
\begin{equation}
\sum\limits_{\alpha} K^{\alpha\alpha}_{\beta_1\beta_2}
(x_{1},x_{2};y_{1},y_{2})
\equiv\delta_{\beta_1\beta_2} K(x_{1},x_{2};y_{1},y_{2})\,.
\end{equation}
In the region (\ref{y-x-condition}),
$0<x_{k}<y_{k}$, the kernel $K$ is given by the following expression
\cite{BFKL-85}
\[
K(x_{1},x_{2};y_{1},y_{2})=\frac{C_{F}}{2\pi^{2}}
\]
\begin{equation}
\times\frac{\delta(y_{1}
-x_{1}-y_{2}+x_{2})}{y_{1}-x_{1}}\left(  1+\frac{x_{1}x_{2}}{y_{1}y_{2}
}\right)  \,.
\end{equation}
Obviously this kernel has the form (\ref{K-alpha})
\[
K(x_{1},x_{2};y_{1},y_{2})=\frac{C_{F}}{2\pi^{2}}\delta(y_{1}-x_{1}
-y_{2}+x_{2})
\]
\begin{equation}
\times\left(  \frac{1}{\sqrt{y_{1}-x_{1}}}\frac{1}{\sqrt{y_{2}-x_{2}}
}+\frac{x_{1}}{y_{1}\sqrt{y_{1}-x_{1}}}\frac{x_{2}}{y_{2}\sqrt{y_{2}-x_{2}}
}\right)  \,.
\end{equation}
This contribution comes from the diagram of Fig.~\ref{Fig-2} with a gluon
playing the role of the intermediate parton.
The two terms in the brackets in the RHS
correspond to two possible polarizations of this gluon.

The above expression for the kernel $K$ is singular at the points $y_{k}=x_{k}$.
The proper treatment of these singularities is described by the following
expression for the convolution of the kernel $K$ with arbitrary functions
$h(x,y)$ \cite{Ji-97,Radyushkin-97,BM-98}
\[
\int\limits_{0}^{\infty}dx_{1}\int\limits_{x_{1}}^{\infty}dy_{1}
\int\limits_{0}^{\infty}dx_{2}\int\limits_{x_{2}}^{\infty}dy_{2}h(x_{1}
,y_{1})h^{\ast}(x_{1},y_{1})
\]
\[
\times K(x_{1},x_{2};y_{1},y_{2})
=\frac{C_{F}}{2\pi^{2}}\lim_{\varepsilon\rightarrow0}\int\limits_{0}^{\infty
}dx_{1}\int\limits_{0}^{\infty}dx_{2}
\left\{  \int\limits_{\varepsilon
}^{\infty}\frac{dz}{z}
\right.
\]
\[
\times
\left[  1+\frac{x_{1}x_{2}}{(x_{1}+z)(x_{2}+z)}\right]
 h(x_{1},x_{1}+z)h^{\ast}(x_{2},x_{2}+z)
\]
\begin{equation}
\left.  +\left(  \frac{3}{2}
-\ln\frac{4x_{1}x_{2}}{\varepsilon^{2}}\right)
h(x_{1},x_{1})h^{\ast}(x_{2},x_{2})\right\}  \,.
\label{K-quark-reg}
\end{equation}
The last term in the RHS proportional to $h(x_{1},x_{1})h^{\ast}(x_{2},x_{2})$
corresponds to the virtual $\delta(y_{1}-x_{1})\delta(y_{2}-x_{2})$
contributions to the evolution kernel $K$. In the limit $\varepsilon
\rightarrow0$ both integral and contact terms are divergent but these two
divergences cancel each other.

\section{Stability of positivity bounds under evolution}
\label{positivity-evolution-appendix}

In this Appendix we present a detailed derivation of the stability of the
positivity bounds (\ref{starting-inequality}) under the
one-loop evolution (\ref{evolution})
to higher normalization points. We shall use compact Dirac
notation for the integral appearing in the LHS of inequality (\ref{starting-inequality})
\[
\langle f|G|f\rangle \equiv \sum\limits_{i_{1}\alpha _{1}i_{2}\alpha
_{2}}\int \frac{dP_{1}^{+}d^{2}P_{1}^{\perp }dx_{1}}{2P_{1}^{+}(2\pi )^{3}}\int
\frac{dP_{2}^{+}d^{2}P_{2}^{\perp }dx_{2}}{2P_{2}^{+}(2\pi )^{3}}
\]
\begin{equation}
\times \theta (x_{1})\theta (x_{2})\,f_{i_{2}\alpha _{2}}^{\ast
}(x_{2},P_{2})f_{i_{1}\alpha
_{1}}(x_{1},P_{1})G_{P_{1}i_{1},P_{2}i_{2}}^{\alpha _{1}\alpha
_{2}}(x_{1},x_{2})
\label{Dirac-notation}
\end{equation}
so that the positivity bound (\ref{starting-inequality}) can be written
as follows
\begin{equation}
\langle f|G|f\rangle \geq 0\,.
\end{equation}

Let us assume that at some normalization point inequality (\ref{starting-inequality})
holds for all functions $f_{i\alpha }(x,P)$. Imagine
that during the evolution to higher $\mu $ this inequality breaks down at some
point $\mu _{0}$ for some function $f^{(0)}$ so that at $\mu >\mu _{0}$\begin{equation}
\langle f^{(0)}|G_{\mu }|f^{(0)}\rangle <0\quad (\mu >\mu _{0})
\label{mu-bad}
\end{equation}but
at $\mu <\mu _{0}$ we still have for all functions $f$\begin{equation}
\langle f|G_{\mu }|f\rangle \geq 0\quad (\mu \leq \mu _{0})\,.
\end{equation}Then
at point $\mu _{0}$ 
\begin{equation}
\langle f^{(0)}|G_{\mu _{0}}|f^{(0)}\rangle =0\,.
\end{equation}The
fate of this ``matrix element'' at $\mu >\mu _{0}$ is determined by the
evolution equation (\ref{evolution})\begin{equation}
\mu \frac{\partial }{\partial \mu }\langle f^{(0)}|G_{\mu }|f^{(0)}\rangle
=\langle f^{(0)}|(K\otimes G_{\mu })|f^{(0)}\rangle \,.
\end{equation}Here
we use short notation $K\otimes G_{\mu }$ for the convolution of the
evolution kernel $K$ with GPD $G$. If one could show that at the point $\mu
_{0}$ we have\begin{equation}
\langle f^{(0)}|(K\otimes G_{\mu _{0}})|f^{(0)}\rangle \geq 0
\end{equation}then
this would guarantee that the evolution from $\mu _{0}$ to higher $\mu $
would restore the positivity of $\langle f^{(0)}|G_{\mu }|f^{(0)}\rangle $.
This would invalidate our assumption (\ref{mu-bad}). Thus in order to prove
the stability of the positivity bounds on GPD under the evolution
to higher normalization points we need the following

\textbf{Statement.} If at some normalization point $\mu _{0}$ for all
functions $f$\begin{equation}
\langle f|G_{\mu _{0}}|f\rangle \geq 0  \label{f-G-f-positive}
\end{equation}and
for some $f^{(0)}$\begin{equation}
\langle f^{(0)}|G_{\mu _{0}}|f^{(0)}\rangle =0  \label{f0-G-f0}
\end{equation}then\begin{equation}
\langle f^{(0)}|(K\otimes G_{\mu _{0}})|f^{(0)}\rangle \geq 0\,.
\label{f0-K-f0-positive}
\end{equation}

To prove this statement we make use of the positivity property of the
evolution kernels (\ref{kernel-positivity}). As it is explained in
Appendix~\ref{positivity-kernels-appendix}
this positivity property is equivalent to the decomposition (\ref{K-beta-decomp})
of the evolution kernel. In our short Dirac-like notations Eq. (\ref{K-beta-decomp})
takes the following form
\begin{equation}
K=\int dz \sum\limits_{\gamma}
|W_{\gamma,z}\rangle \langle W_{\gamma,z}|
\label{K-naive-decomposition}
\end{equation}
Inserting representation (\ref{K-naive-decomposition})
for $K$ into the LHS of inequality (\ref{f0-K-f0-positive})
we find
\[
\langle f^{(0)}|(K\otimes G_{\mu _{0}})|f^{(0)}\rangle 
\]\begin{equation}
=\int dz \sum\limits_{\gamma}\langle f^{(0)}\otimes W _{\gamma,z }|G_{\mu
_{0}}|f^{(0)}\otimes W _{\gamma,z }\rangle \geq 0\,.
\label{K-beta-positive}
\end{equation}
The RHS is positive according to Eq.~(\ref{f-G-f-positive}). This completes the
derivation of the inequality (\ref{f0-K-f0-positive}).

The above proof of the stability of the positivity bounds (\ref{starting-inequality})
under the evolution (\ref{evolution}) ignored the
problem of the negative terms proportional to $\delta (x_{1}-y_{1})\delta
(x_{2}-y_{2})$ which are present in the evolution kernel $K$. Because of
these terms it is allowed to use inequality (\ref{kernel-positivity}) only
for functions $h(x,y)$ which vanish at $x=y$. As a result instead of
Eq.~(\ref{K-naive-decomposition})
one has to work with the following representation
for the evolution kernel\begin{equation}
K=\lim_{\varepsilon \rightarrow +0}\left( K_{\varepsilon
}^{(1)}+K_{\varepsilon }^{(2)}\right)   \label{K-decomposition}
\end{equation}where
the first term is given by the $\varepsilon $ regularized integral decomposition
(\ref{K-naive-decomposition})
\begin{equation}
K_{\varepsilon }^{(1)}(x_{1},x_{2},y_{1},y_{2})=\int\limits_{\varepsilon
}^{\infty }dz \sum\limits_{\gamma}W _{\gamma,z }(x_{1},y_{1})W _{\gamma,z
}^{\ast }(x_{2},y_{2})  \label{K-decomposition-1}
\end{equation}and the
second piece
\begin{equation}
K_{\varepsilon }^{(2)}(x_{1},x_{2},y_{1},y_{2})=C(\varepsilon
,x_{1},x_{2})\delta (x_{1}-y_{1})\delta (x_{2}-y_{2})  \label{K-2-C}
\end{equation}comes
from the virtual contribution to the evolution kernel which is
proportional to $\delta (x_{1}-y_{1})\delta (x_{2}-y_{2})$ with a divergent
coefficient $C(\varepsilon ,x_{1},x_{2})$. This coefficient is regularized
by the same small parameter $\varepsilon $ so that\begin{equation}
C(\varepsilon ,x_{1},x_{2})=c_{1}+c_{2}\ln \varepsilon
+F_{1}(x_{1})+F_{2}(x_{2})\,.  \label{C-def}
\end{equation}The
singular contribution $\ln \varepsilon $ compensates the divergence of
the $z $ integral in the RHS of Eq.~(\ref{K-decomposition-1}).
Obviously representation (\ref{K-decomposition})
guarantees that $K$ obeys inequality (\ref{kernel-positivity})
for arbitrary functions $h(x,y)$ vanishing at $x=y$.
An example of the general structure (\ref{C-def}) can be seen in the explicit
expression (\ref{K-quark-reg}) for the quark evolution kernel where the term
$\ln((4x_1x_2)/\varepsilon^2)$ represents all terms appearing in the rhs of
(\ref{C-def}).

Now we have to show that the kernel $K$ (\ref{K-decomposition}) still has
the property (\ref{f0-K-f0-positive}). For the first term in the RHS of Eq. (\ref{K-decomposition})
this can be done in the same way as in the inequality (\ref{K-beta-positive}):
\begin{equation}
\langle f^{(0)}|(K_{\varepsilon }^{(1)}\otimes G_{\mu _{0}})|f^{(0)}\rangle
\geq 0\,.  \label{f0-K1-f0-positive}
\end{equation}

Next let us show that
\begin{equation}
\langle f^{(0)}|(K_{\varepsilon }^{(2)}\otimes G_{\mu _{0}})|f^{(0)}\rangle
=0\,.
\label{f0-K2-f0}
\end{equation}
According
to Eqs. (\ref{K-2-C}), (\ref{C-def}) we have\[
\langle f^{(0)}|(K_{\varepsilon }^{(2)}\otimes G_{\mu _{0}})|f^{(0)}\rangle 
=\langle f^{(0)}| [(c_{1}+c_{2}\ln \varepsilon)G_{\mu _{0}}
\]
\begin{equation}
 +G_{\mu _{0}}F_{1}+F_{2}G_{\mu
_{0}}]|f^{(0)}\rangle \,.  \label{K-2-c-F}
\end{equation}
Note that $|f^{(0)}\rangle $ is associated with $f^{(0)}(x_{1},y_{1})$
whereas $\langle f^{(0)}|$ corresponds to $f^{(0)}(x_{2},y_{2})$
in Eq.~(\ref{Dirac-notation}). This
explains the ordering of ``operators'' $F_{1}$ and $F_{2}$ associated with
functions $F_{1}(x_{1})$ and $F_{2}(x_{2})$ respectively in the RHS of
Eq.~(\ref{K-2-c-F}).
Taking into account Eq. (\ref{f0-G-f0}) we get rid of the $c_{1}+c_{2}\ln
\varepsilon $ contribution:\[
\langle f^{(0)}|(K_{\varepsilon }^{(2)}\otimes G_{\mu _{0}})|f^{(0)}\rangle 
\]
\begin{equation}
=\langle f^{(0)}|(G_{\mu _{0}}F_{1}+F_{2}G_{\mu _{0}})|f^{(0)}\rangle \,.
\label{K2-G-F}
\end{equation}
Since property of $G_{\mu _{0}}$  (\ref{f-G-f-positive}) holds for any $f$ we
conclude from Eq. (\ref{f0-G-f0}) that
\begin{equation}
G_{\mu _{0}}|f^{(0)}\rangle =0\,,\quad \langle f^{(0)}|G_{\mu _{0}}=0\,.
\end{equation}
This means that
\begin{equation}
\langle f^{(0)}|(G_{\mu _{0}}F_{1}+F_{2}G_{\mu _{0}})|f^{(0)}\rangle =0\,.
\end{equation}
Inserting this result into Eq. (\ref{K2-G-F}) we obtain Eq.~(\ref{f0-K2-f0}).
Combining Eqs.~(\ref{f0-K1-f0-positive}), (\ref{f0-K2-f0})
we derive inequality (\ref{f0-K-f0-positive}),
which means that the virtual terms in the
evolution kernel do not violate the stability of the positivity bounds on
parton distributions under the evolution to higher normalization points.

\section{Derivation of positivity bounds in impact parameter representation}
\label{ineq-deriv-appendix}

In this Appendix we show how inequality (\ref{starting-inequality})
containing a multidimensional integral can be reduced to the relatively
simple form (\ref{main-ineq}).

Using representation (\ref{standard-GPD-F}) for GPD $G$ we find from
inequality (\ref{starting-inequality})
\[
\int\limits_{0}^{\infty }dk^+ _{1}\int\limits_{0}^{\infty }\frac{
dP_{1}^{+}}{P_{1}^{+}}\int d^{2}P_{1}^{\perp }\,\int\limits_{0}^{\infty
}dk^+ _{2}\int\limits_{0}^{\infty }\frac{dP_{2}^{+}}{P_{2}^{+}}\int
d^{2}P_{2}^{\perp }
\]
\[
\times \delta \left[ (P_{2}^{+}-k^+ _{2})-(P_{1}^{+}-k^+ _{1})\right]
\]
\[
\times F_{\sigma _{2}\sigma _{1}}\left[ \frac{k^+ _{1}+k^+ _{2}}{
P_{1}^{+}+P_{2}^{+}},\frac{P_{1}^{+}-P_{2}^{+}}{P_{1}^{+}+P_{2}^{+}},\frac{
2(P_{1}^{+}P_{2}^{\perp }-P_{2}^{+}P_{1}^{\perp })}{P_{1}^{+}+P_{2}^{+}}
\right] 
\]
\begin{equation}
\times f_{\sigma _{2}}^{\ast }(k^+ _{2},P_{2}^{+},P_{2}^{\perp
})f_{\sigma _{1}}(k^+ _{1},P_{1}^{+},P_{1}^{\perp })\geq 0\,.
\label{ineq-1}
\end{equation}
Taking functions $f_{\sigma }$ in the factorized form
\begin{equation}
f_{\sigma }(k^+ ,P^{+},P^{\perp })=d_{\sigma }(k^+ ,P^{+})\eta \left( 
\frac{P^{\perp }}{P^{+}}\right) 
\end{equation}
we derive from inequality (\ref{ineq-1})
\[
\int\limits_{0}^{\infty }dk^+ _{1}\int\limits_{0}^{\infty
}dP_{1}^{+}\int\limits_{0}^{\infty }dk^+ _{2}\int\limits_{0}^{\infty
}dP_{2}^{+}d_{\sigma _{2}}^{\ast }(k^+ _{2},P_{2}^{+})d_{\sigma
_{1}}(k^+ _{1},P_{1}^{+})
\]
\[
\times \tilde{F}_{\sigma _{2}\sigma _{1}}\left[ \frac{k^+ _{1}+k^+ _{2}
}{P_{1}^{+}+P_{2}^{+}},\frac{P_{1}^{+}-P_{2}^{+}}{P_{1}^{+}+P_{2}^{+}},\frac{
P_{1}^{+}+P_{2}^{+}}{2P_{1}^{+}P_{2}^{+}}b^{\perp }\right] 
\]
\begin{equation}
\times
\delta \left[ (P_{2}^{+}-k^+ _{2})-(P_{1}^{+}-k^+ _{1})\right]
\left( \frac{P_{1}^{+}+P_{2}^{+}}{2P_{1}^{+}P_{2}^{+}}\right) ^{2}\geq 0
\label{ineq-2}
\end{equation}
where $\tilde{F}_{\sigma _{2}\sigma _{1}}$ is defined by Eq. (\ref{F-Fourier}).
This inequality should hold for any value of $b^{\perp }$ and for any
function $d_{\sigma }(k^+ ,P^{+})$. The next step is to take
\begin{equation}
d_{\sigma }(k^+ ,P^{+})=g(P^{+}-k^+ )h_{\sigma }(P^{+})\,.
\end{equation}
In the limit
\begin{equation}
\left| g(u)\right| ^{2}\rightarrow \delta (u-\nu )
\end{equation}
we find from inequality (\ref{ineq-2})
\[
\int\limits_{-1}^{1}d\xi \int\limits_{|\xi |}^{1}dx\frac{1}{(1-x)(1-\xi
^{2})^{2}}
h_{\sigma _{2}}^{\ast }\left( \nu \frac{1-\xi }{1-x}\right) 
\]
\begin{equation}
\times
h_{\sigma_{1}}\left( \nu \frac{1+\xi }{1-x}\right) 
 \tilde{F}_{\sigma _{2}\sigma _{1}}\left( x,\xi ,\frac{(1-x)b^{\perp }
}{\nu (1-\xi ^{2})}\right) \geq 0\,.  \label{ineq-3}
\end{equation}
Since function $h_{\sigma }$ and parameter $b^{\perp }$ are arbitrary we can
set $\nu =1$ without losing generality. Taking $h_{\sigma
}(z)=z^{2}p_{\sigma }(z^{-1})$ we obtain inequality (\ref{main-ineq}).

\end{document}